\documentclass[]{cjp}

\usepackage{graphicx}

\def\eion{{(e~+~ion)}\ }
\def\ra{{$\rightarrow$}\ }
\def\fexvii{{\rm Fe~\sc xvii}\ }
\def\fexviii{{\rm Fe~\sc xviii}\ }
\def\fexix{{\rm Fe~\sc xix}\ }
\def\fexx{{\rm Fe~\sc xx}\ }

\def\en{{$n$\ }}
\def\el{{$l$\ }}
\def\ii{{$i$\ }}
\def\dne{{$N_e$\ }}
\def\om{{$\omega$\ }}
\def\sig{{$\sigma$\ }}
\def\gam{{$\Gamma$\ }}
\def\gamc{{$\Gamma_c$\ }}
\def\gamd{{$\Gamma_d$\ }}
\def\gams{{$\Gamma_s$\ }}

\newcommand{\be}{\begin{equation}}
\newcommand{\ee}{\end{equation}}

\volyear{}[]{}
\received{}
\accepted{}

\begin{document}

\title{Plasma Effects on Resonant Phenomena}
\author{Anil Pradhan} 
\address{Department of Astronomy, Chemical Physics Program, The Ohio State University, Columbus, OH 
43210, USA}
%\address{Department of Astronomy, The Ohio State University, Columbus, OH 
%43210, USA}
%\address{Institution,
%Somewhere Else, CA 94550, Country}
%\AddressNote{also at this Institution}
%\address{Department of Astronomy, The Ohio State University, Columbus, OH 
%43210, USA}
%\AddressNote{also at this Institution}

%\shortauthor{A. Someone and Someone Else}
\shortauthor{Pradhan}

\maketitle

\begin{abstract}

The effect of autoionizing 
resonances in atomic systems and processes is reviewed.
Theoretical framework for treating resonances in the coupled channel
approximation using the R-matrix method, as well as approximations
related to plasma applications are described. The former
entails large-scale atomic computations, and the latter is based on
a new method for including collisional, Stark, thermal and other
broadening mechanisms. We focus particularly on the
problem of opacities calculations in high-energy-density (HED) plasmas such as
stellar interiors and inertial confinement fusion devices. The treatment
is generally relevant to radiative and collisional processes as the cross
sections become energy-temperature-density dependent. While the
computational difficulty increases considerably, the reaction rates are 
significantly affected. The related issue of the Boltzmann-Saha 
equation-of-state and its variants in local-thermodynamic-equilibrium
(LTE) is also explored as the intermediary between atomic data
on the one hand and plasma environments on the other.

\end{abstract}

\medskip


\date{today}

\section{Introduction}

 Resonant phenomena are ubiquitous in physics and stem from correlation
effects. In AMO physics autoionizing resonances are due to electron
correlation effects among bound and continuum states. They manifest
themselves prominently in cross sections of various atomic and molecular
processes in laboratory and astrophysical plasma sources. The shapes,
magnitudes, and extent of resonances determines associated rates for spectral
formation and experimental and observed spectra. The well-known
Fano profile \cite{fano,rau,fr}
is widely used to analyze isolated resonance structures,
with parameters that may be compared with theoretical calculations. 

On the other hand, overlapping infinite series of autoionizing
resonances converging on to large numbers of excited levels are also of great
importance. In general, they are not amenable to analytic 
formulation and require
computationally intensive coupled channel calculations, the most
powerful of which is the R-matrix method by P.G. Burke and collaborators
\cite{b11}.\footnote{Following the original suggestion by U. Fano about
the R-matrix theory of nuclear reactions
by A.M. Lane and R.G. Thomas \cite{lt}, P.G. Burke
introduced it for atomic and molecular processes, as described
in \cite{b11}. Later, M.J. Seaton and collaborators adapted
the R-matrix method for large-scale calculations required for the
Opacity Project \cite{op}.}
Whereas the R-matrix method has long been utilized for a
variety of atomic processes and applications, I focus on a
large-scale application to plasma opacities that is of immense importance in
astrophysics and nuclear fusion sources.
R-matrix atomic calculations including resonant phenomena in
an {\it ab initio} manner, and taking account of plasma effects, are
exceedingly difficult but no longer computationally intractable for
complex atoms and ions \cite{p1,p2,p3,p4}.

Fano derived his general formula in the early days of experimental and
theoretical autoionization studies, but before the
close coupling codes were developed to delineate resonance
profiles precisely (viz.\cite{aas}). However, HED plsama applications
require not only a complete description of the bound-free resonant continua, 
but also interface with the equation-of-state and
resonance broadening and level-dissolution effects as function
of temperature and density \cite{p3,eos}.
{\it But in fact, the intrinsically quantum mechanical autoionization is the
dominant broadening mechanism when extrinsic plasma effects are negligible or
small, and autoionization widths are much bigger than line widths due to
radiation damping, and/or Stark, thermal, or collisional broadening.} 
In particular, the characteristically asymmetric Seaton
resonances due to photoexciation-of-core (PEC) in photoionization cross
sections extend over large energy ranges of up to hundreds of eV, and
can only be described by coupled channel correlations \cite{ys,aas,p2}. 
Furthermore, the Seaton PECs retain their resonance profiles even as
other narrower Rydberg series of resonances dissolve into the continuum with
increasing plasma densities.
For these reasons, R-matrix coupled channel calculations 
would be necessary in general for
all atomic systems under any plasma conditions for high precision, and
to be preferred over simpler approximations based on the distorted wave
method (DW) or its variants that do not include autoionizing resonance
structures (discussed later).

Opacity determines the light we see or detect. All radiation-matter
interactions need to be considered in order to determine the opacity of
a given plasma source or medium.
Primary physical processes contributing to opacity are:
\begin{equation}
 \kappa_{ijk}(\nu) = \sum_k A_k \sum_j F_j \sum_{i,i'}
[\kappa_{bb(}(i,i';\nu) +
\kappa_{bf}(i,\epsilon_{i'};\nu) + \kappa_{ff} (\epsilon_i,
\epsilon'{_i'};
\nu) + \kappa_{sc} (\nu)]\ .
\label{eq:k}
\end{equation}

The first two are the dominant processes;
 $bb$ refers to bound-bound and $bf$ to bound-free transitions or
photoionization. The other two, free-free transitions and photon
scattering, are generally much smaller and may be treated by simple
approximations (viz. \cite{op,aas}).
In Eq.~\ref{eq:k}, $A_k$ is the abundance of an element $k$, its ionization
fraction $F_j$ at a given temperature-density, $i,i'$ are 
initial bound and final bound/continuum states,
and $\epsilon$ represents the free electron energy.

\section{Coupled channel R-matrix method and atomic-plasma effects}

 The state-of-the-art R-matrix (RM) method provides a powerful
computational tool to implement the general coupled-channel theoretical
framework. An atomic systems is represented by a N-electron core or target ion
wavefunction $\chi_i$ coupled with an $(N+1)^{th}$ free 
electron wavefunction $\theta_i$ in a bound or continuum state of the
\eion system. The total \eion wavefunction is then a
quantum superposition expressed as
\begin{equation}
\Psi(E) = \mathcal{A} \sum_{i}(E_i) \chi_{i}\theta_{i}(k_i^2) + \sum_{j} c_{j}
\Phi_{j}(E_j).
\label{eq:psi}
\end{equation}

When the free electron kinetic energy $k_i^2 > 0$, 
the first sum on the RHS of
Eq.~\ref{eq:psi} represents a coupled channel system for electron-ion
scattering or half-scattering photoionization process \cite{fr}. 
Each channel is defined by the spin-orbital quantum symmetries
$(S_iL_iJ_i) \ \ell_is_i [SLJ]$. 
Rydberg series of resonances arise from photoexcitation of bound energy 
levels into \eion
continua comprising of coupled thresholds of the core ion. A
particular type of resonances due to photoexcitation-of-core (PEC) or
Seaton resonances are due to strong dipole transitions in the core
ion where the continuum electron in a Rydberg level 
remains a 'spectator' \cite{ys,aas,p2}. The Seaton PEC
resonances constitute the detailed balance 
inverse of the dielectronic recombination process, wherein
an (photo-)excited core ion undergoes radiative decay by emission of
photons redward of the core transition wavelength. 
The huge
Seaton resonances dominate bound-free opacity under all plasma
conditions.

If the first summation on the RHS of Eq.~\ref{eq:psi} is neglected then
the coupling effects in the \eion wavefunction are excluded, yielding
the distorted wave (DW) approximation that does not include autoionizing
resonances in an {\it ab initio} manner as the RM method.
Owing to its simplicity, the DW method has been employed in existing
opacity models. While resonant phenomena are not included in the DW
calculations, that contribution may be included perturbatively
by considering autoionizing resonances on par with lines as bound-bound
transitions and employing line broadening theory for plasma broadening.
However, the detailed auotionization shapes over
extended energy ranges and their precise effect on atomic rates
is not taken into account in DW models.

The Opacity Project (OP \cite{op}) was originally
developed to implement the RM method but was computationally
intractable for most complex atoms. In particular, it was found that
inner-shell transitions from closed electronic shells into outer open
subshells implies a large number of channels to be included in
Eq.~\ref{eq:psi}.
In recent years, a renewed and extended version of OP has been initiated
\cite{np16} 
and is now in progress for improved opacities with higher
accuracy \cite{p1,p2,p3,p4}. 

For opacity calculations, 
the transition matrix elements are obtained with dipole operator ${\bf D}
=\sum_i{\bf r}_i$, where the sum is over all electrons, which
gives the generalized line strength as,
\begin{equation}
{\bf S= \left|\left\langle{\mit\Psi}_f
 \vert\sum_{j=1}^{N} r_j\vert
 {\mit\Psi}_i\right\rangle\right|^2 
\label{eq:ls}}
\end{equation}
The oscillator strength ($f_{ij}$), radiative decay rate ($A_{ji}$),
photoionization cross section ($\sigma_{PI}$), and mass attenuation
coefficient then can be expressed as follows
\begin{equation}
f_{ij} = \left [{E_{ji}\over {3g_i}}\right ]S,~~ 
A_{ji}(sec^{-1}) = \left [0.8032\times 10^{10}{E_{ji}^3\over
{3g_j}}\right ]S,
~
\sigma_{PI}(K\alpha,\nu)=\frac{4\pi^{2}a_{o}^{2}\alpha}{3}\frac{E_{ij}}
{g_{k}} S 
%=4\pi^{2}a_{o}^{2}\alpha f_{ij}
\label{eq:fp}
\end{equation}

The Breit-Pauli R-matrix (BPRM) method incorporates
relativistic effects using the Breit-Pauli (BP)
Hamiltonian for the \eion system
 in intermediate coupling, with a pair-coupling
scheme $S_iL_l(J_i)l_i(K_i)s_i (SLJ\pi)$, whereby
states $S_iL_i$ is split into fine-structure levels $S_iL_iJ_i$, and
$SLJ\pi$ is the total spin-orbital symmetry.
Consequently, the number of
channels becomes several times larger than the corresponding $LS$
coupling case. A considerable body of work with the BPRM codes has been carried
out under the follow-on project to OP, the Iron Project \cite{ip}. 
The IP work is based on BPRM codes and archived in the
large amount of radiative and collisional data in databases
NORAD \cite{norad} and OP/IP database Topbase \cite{top}.

\section{Plasma environment and approximations}

 The practical limitation of the RM method for plasma applications
is evident from
Eq.~\ref{eq:psi}. Computational constraints imply that only a 
finite and usually small 
number of excited core ion states and resulting channels may be
explicitly included in the wavefunction expansion.
This has implications
in high-energy-density (HED) plasmas such as in stellar interiors or fusion
devices, wherein a large number of excited states exist and
differentially perturbed. The temperature regime may be in excess of
$10^6-10^7$ K, with electron densities up to $10^{27}$ cm$^{-3}$.
Among the largest R-matrix calculations carried out thus far 
are the recent ones for Fe ions
\fexvii, \fexviii and \fexix that constitute $\sim$85\% of iron opacity at the 
boundary between the solar radiative and convection zones at radius 
$R_{\odot}$ = 0.713$\pm$0.001, where T=$2\times10^6$K and electron density
$N_e = 10^{23}$ cm$^{-3}$ \cite{p1}. The number of core levels included
in the R-matrix \eion wavefunction expansion were 218 levels of \fexviii for
the (e+\fexviii) \ra \fexvii bound and continuum states, 276 levels of
\fexix for the (e+\fexix) \ra \fexviii, and 99 LS terms of \fexx for
(e+\fexx) \ra \fexix \cite{p2}.

\subsection{Equation-of-state}
 
 In addition to the theoretical limit of the RM method, an obvious limit
is imposed by perturbations on atoms by the 
plasma environment depending on the specific temperature-density.
That manifests itself via the equation-of-state that determines the
 atomic ionization state and level populations. The generally employed
approximation is to assume local-thermodynamic-equilibrium (LTE), as
defined by the Saha-Boltzmann equations. From an atomic-plasma physics
point of view, a widely employed formulation (such as in OP) is the 
Mihalas-Hummer-D\"{a}ppen (MHD) equation of state in the so-called
"chemical picture"\cite{mhd}. It is based on the
concept of {\it occupational probability} $w$ of an excited level being
occupied at a given temperature and density such that the level
population is

\be N_{ij} = \frac{N_j  g_{ij} w_{ij} e^{-E_{ij}/kT}}{U_j}, \label{eq:lp}
\ee

 where $w_{ij}$ are the occupation probabilities of levels $i$ in
  ionization state $j$, and $U_j$ is the atomic internal partition
function. The occupation probabilities do not have a sharp
  cut-off, but approach zero for
  high-\en as they are dissolved due to plasma interactions.
 The partition function is re-defined as

 \be U_j = \sum_i g_{ij} w_{ij} e^{(-E_{ij}/kT)}. \label{eq:pf} \ee

$E_{ij}$ is the excitation energy of level $i$, $g_{ij}$ its statistical
weight and $T$ the temperature. The $w_{ij}$ are obtained by
free-energy minimization, and taking into
account Stark ionization due to plasma microfields \cite{mhd}.
Hence, the exact form of the equation-of-state numerically determines
how many and how much the excited states of an atom contribute to
opacity and radiation transport.

\subsection{Broadening of autoionizing resonances}

 Unlike line broadening {\it intrinsic} autoionization (AI) decay rates
are much larger relative to radiative rates. Therefore, one expects 
resonances to broaden, smear out, and dissolve into the continuum much more
than lines when subjected to {\it extrinsic} 
HED plasma environments. Also, unlike line broadening for which
theoretical formulations are well developed and long employed, there is
no {\it ab initio} and general treatment for AI broadening.
Even for line broadening the most elaborate methods are 
are precise only for hydrogenic and
simple atomic systems and several approximations are necessary to apply
those to complex atoms in realistic sources \cite{op}. 

 Recently, a general theoretical and computational 
formalism has been introduced for AI resonance broadening
\cite{p3}. Analogous to line broadening, the physical mechanisms
considered are: electron collisions (pressure broadening), ion
microfields (Stark broadening), Doppler effect (thermal broadening),
and free-free transitions. It has been shown that {\it extrinsic} plasma
effects redistribute and shift resonance strengths, even as the broad
{\it intrinsic} asymmetries of resonance profiles are discernible.
Furthermore, while the shapes, magnitudes and extent of
resonances are affected,
the total integrated resonance oscillator strengths are conserved
and manifest in the bound-free continua independent of temperature and density.
The energy-temperature-density dependent
cross sections would elicit and introduce physical features in resonant
processes in photoionization, \eion excitation and recombination.
The method should be generally applicable to
atomic species in high-energy-density (HED)
sources such as fusion plasmas and stellar interiors.

 Whereas the main
broadening mechanisms in AI broadening are physically similar to line
broadening, their
theoretical and computational treatment is quite different. Superimposed
on intrinsic AI broadening in atomic cross sections
the extent of resonances owing to extrinsic plasma
effects renders much of the line broadening theory inapplicable,
particularly for multi-electron systems. The {\it unbroadened}
AI resonances themselves vary by orders of magnitude in width, shapes
and heights, and
incorporate two types: large features due to
photoexcitation-of-core (PEC) below thresholds corresponding
to dipole core transitions \cite{ys}, and infinite
Rydberg series of
resonances converging on to each excited core level of the \eion
system. The generally employed Voigt line profiles obtained
by convolution of a Lorentzian function for radiative and
collisional broadening, and a Gaussian function for Doppler or thermal
broadening, are found to be practically inapplicable
for AI broadening. Numerically, the Voigt kernel is ill-conditioned
since the collisional-to-Doppler width ratio \gamc/\gamd
varies over a far wider range
for resonances than lines and therefore unconstrained {\it a priori}.

 The physical processes for broadening of AI resonances differ from
lines qualitatively and quantitatively.
However, line broadening processes and formulae may be generalized
to develop a
theoretical treatment and computational algorithm outlined herein
(details to be presented elsewhere).
The convolved bound-free photoionization cross section of level \ii
may be written as:

\be \sigma_i(\omega) = \int \tilde{\sigma}(\omega') \phi
(\omega',\omega) d\omega', \label{eq:conv} \ee

where \sig and $\tilde{\sigma}$ are the cross sections with
plasma-broadened
and unbroadened AI resonance structures, \om is the photon energy
(Rydberg atomic units are used throughout), and $\phi (\omega',\omega)$
 is the normalized
Lorentzian profile factor in terms of the {\it total} width \gam due to
all
AI broadening processes included:

\be \phi (\omega',\omega) = \frac{\Gamma(\omega)/\pi}{x^2+\Gamma^2},
\label{eq:prof} \ee

where $x \equiv \omega-\omega'$. The crucial difference with line
broadening is
that AI resonances in the \eion system correspond to and are due to
quantum mechanical interference between discretized continua
defined by excited core ion levels in a multitude of channels. The
coupled channel (CC) approximation, such as implemented by the R-matrix
(RM) method
(viz. \cite{b11,op,aas}), accounts for AI resonances in an \eion
system
with generally asymmetric profiles (unlike line profiles that are
usually symmetric).
Given $N$ core ion levels corresponding to resonance
structures,

\be \sigma(\omega) = \sum_i^N \left[ \int \tilde{\sigma}(\omega')
\left[ \frac{\Gamma_i(\omega)/\pi}{x^2 +
\Gamma_i^(\omega)}\right] d \omega' \right]
. \label{eq:sig} \ee

 With $x \equiv \omega' - \omega $, the summation is over all excited
thresholds $E_i$ included in
the $N$-level CC or RM wavefunction expansion, and corresponding
to total damping width $\Gamma_i$ due to all broadening processes.
The profile $\phi(\omega',\omega)$ is centered at each
continuum energy $\omega$, convolved over the variable $\omega'$ and
relative to each excited core ion threshold \ii.
In the present formulation we associate the energy to the effective
quantum number relative to each threshold $\omega' \rightarrow \nu_i$ to
write the total width as:

\begin{eqnarray}
\Gamma_i(\omega,\nu,T,N_e) &  = & \Gamma_c(i,\nu,\nu_c)+
\Gamma_s(\nu_i,\nu_s^*)\\
 & + &  \Gamma_d(A,\omega) + \Gamma_f(f-f;\nu_i,\nu_i'), \nonumber 
\label{eq:gam}
\end{eqnarray}

pertaining to
collisional $\Gamma_c$, Stark $\Gamma_s$, Doppler $\Gamma_d$, and
free-free transition $\Gamma_f$ widths
respectively, with additional parameters as defined below.
Without loss of generality we assume a Lorentizan profile
factor that describes collisional-ion broadening which dominates in
HED plasmas. We assume this approximation to be valid since
collisional profile wings extend much wider as $x^{-2}$, compared to
the shorter range $exp(-x^2)$ for thermal Doppler, and $x^{-5/2}$ for
Stark broadening. In evaluating Eq.~(10) from Eq.~\ref{eq:sig}
the limits $\mp \infty$ are
replaced by $\mp \Gamma_i/\sqrt{\delta}$; $\delta$ is
chosen to ensure the Lorentzian profile energy range
for accurate normalization (see Eq.~\ref{eq:pnorm}).
Convolution by evaluation of Eqs.~(\ref{eq:conv},\ref{eq:sig}) 
is carried out for each
energy $\omega$ throughout the tabulated mesh of energies used to
delineate all AI resonance structures, for each cross section,
and each core ion threshold. We employ the following expressions for
computations:

\be \Gamma_c(i,\nu) \  = \ 5 \left( \frac{\pi}{kT} \right)^{1/2}
 a_o^3 N_e G(T,z,\nu_i) (\nu_i^4/z^2), \label{eq:gamc} \ee

where T, \dne, $z$, and $A$ are the temperature, electron density, ion
charge and atomic weight respectively, and $\nu_i$ is the effective
quantum
number relative to each core ion threshold \ii: $\omega \equiv E =
E_i-\nu_i^2/z^2$ is a continuous variable. The Gaunt factor
$G(T,z,\nu_i) = \sqrt 3/\pi [1/2+ln(\nu_i kT/z)]$.
A factor
$(n_x/n_g)^4$ is introduced for $\Gamma_c$
to allow for doubly excited AI levels with excited core
levels $n_x$ relative to the ground configuration $n_g$
(e.g. for \fexviii
$n_x=3,4$ relative to the ground configuration $n_g=2$).
A treatment of the Stark effect for complex
systems entails two approaches, one where both electron and ion
perturbations are combined (viz. \cite{dk87}), or separately (viz.
\cite{op,p3}) employed herein. Excited Rydberg levels are nearly
hydrogenic
and ion perturbations are the main broadening effect, though collisional
broadening competes significantly increasing with density
as well as $\nu_i^4$ (Eq.~5).
The total Stark width of a given
\en-complex is $\approx (3F/z)n^2$, where F is the plasma electric
microfield.
Assuming the dominant ion perturbers to be protons and density
equal to electrons, \dne=$N_p$, we take $F=[(4/3)
\pi a_o^3 N_e)]^{2/3}$, as employed in the Mihalas-Hummer-D\"{a}ppen
equation-of-state formulation \cite{mhd}.

\be \Gamma_s(\nu_i,\nu_s^*) =
[(4/3)\pi a_o^3 N_e]^{2/3} \nu_i^2. \label{eq:gams} \ee

In addition, in employing Eq. (6) a Stark ionization parameter
$\nu_s^* = 1.2\times 10^3 N_e^{-2/15}z^{3/5}$ is introduced such
that AI resonances may be considered fully dissolved into the continuum
for $\nu_i > \nu_s^*$ (analogous
to the Inglis-Teller series limit \cite{it,mhd}).
Calculations are carried out with and without
$\nu_s^*$ as shown later in Table~1. The Doppler width is:

\be \Gamma_d (A,T,\omega) = 4.2858 \times 10^{-7} \sqrt(T/A),
\label{eq:dop} \ee

where $\omega$ is {\em not} the usual line center but taken to be each
AI resonance energy. The last term $\Gamma_f$ in Eq. (5) accounts for
free-free
transitions among autoionizing levels with $\nu_i,\nu_i'$ such that

\be X_i + e(E_i,\nu_i) \longrightarrow X_i' + e'(E_i',\nu_i').
\label{eq:ff} \ee

The large number of free-free transition probabilities for $+ve$ energy
AI
levels $E_i,E_i' > 0$ may be computed using RM or atomic structure
codes.

 Whereas Eq.\ref{eq:sig} has an analytical solution in terms of
$tan^{-1}(x/\Gamma)/\Gamma$ evaluated at limiting values of $x
\rightarrow \mp \Gamma/\sqrt\delta$, its evaluation
for practical applications entails piece-wise integration across
multiple energy ranges spanning many excited thresholds and different
boundary
conditions. For example, the total width $\Gamma$ is very large at
high densities and the Lorentzian profile may be incomplete
above the ionization threshold and therefore not properly normalized.
We obtain the necessary redward left-wing correction for partial
re-normalization as

\be
\lim_{a \rightarrow - \Gamma/2\sqrt\delta}
\int_a^{+\Gamma/\sqrt\delta} \phi(\omega,\omega') d\omega' =
\left[ \frac{1}{4} - \frac{tan^{-1}(\frac{a}{\Gamma/2\sqrt\delta})}{\pi}
\right], \label{eq:pnorm}
\ee

where $a$ is the lower energy range up to the ionization threshold,
reaching
the maximum value $-\Gamma/2\sqrt\delta$.
\section{Results and discussion}
\begin{figure*}
\begin{center}
%\resizebox{150mm}{!}
%{ \includegraphics{pxbpfe17-gd}}
%{ \includegraphics{f1}}
\includegraphics[width=\columnwidth,keepaspectratio]{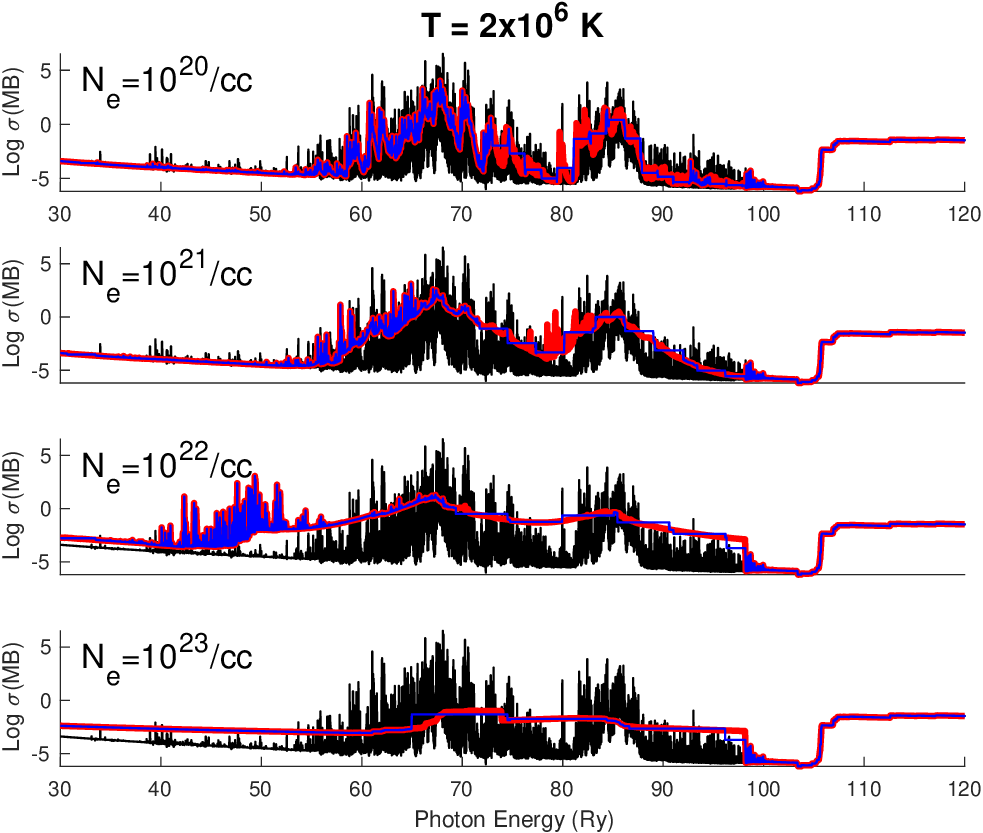}
\end{center}
\vskip 0.25in
\caption{Plasma broadened photoionization cross sections
for $\hbar \omega + \fexviii \rightarrow e~+~\fexix $ of the bound
level $2s^22p^4 \ [^3P^e_0] \  5s  (^2P_{1/2})$ (ionization energy 13.79
Ry), along the isotherm
$T=2 \times 10^6$K and electron densities $N_e=10^{20,21,22,23}$/cc: black
---
unbroadened, red --- broadened, blue --- broadened with Stark
ionization cut-off $\nu_s^*$ (Table 1). Rydberg series of
AI resonance complexes with $\nu_i \leq 10$ belonging to 276 excited
\fexix levels broaden and shift with increasing density, also
 resulting in continuum raising and threshold lowering.
The two large features around 68 RY and 85 Ry are combinations of Seaton
PEC and Rydberg series of resonances.
\label{fig:f1}}
\end{figure*}

All atomic cross sections with resonant phenomena
are modified by the plasma environment. An exemplar from large-scale
opacity calculations \cite{p1,p2,p3,p4} is presented in
Fig.~\ref{fig:f1}.
The complexity and magnitude of computations is demonstrated
for the (e~+~\fexix) $\longrightarrow$ \fexviii system in an highly
excited level $2s^22p^4 \ [^3P^e_0] \  5s  (^2P_{1/2})$ with ionization
energy = 13.79 Ry, compared to the ground level $2s^22p^5 \
^2P^o_{1/2})$ = 98.9 Ry. We utilize new results from an extensive
BPRM calculation
with 276-levels dominated by $n=2,3,4$ levels of the core
ion \fexix \cite{p2}, resulting in 1,601 bound levels of \fexviii with
configurations up to 
$\ n \leq 10, \ \ell \leq 9, \ J \leq 12$). 
 Rydberg series of
AI resonances correspond to $(S_iL_iJ_i) \ n \ell, \ n \leq 10, \ell
\leq 9$, with effective quantum number defined as a continuous variable
$\nu_i = z/\sqrt(E_i-E) \ (E>0)$, up to the
highest
$276^{th}$ \fexix core level.
 AI resonances are resolved for all
cross sections at $\sim$45,000 photon energies \cite{p2,p4}.

Fig.~\ref{fig:f1} displays detailed results for plasma broadened and unbroadened
photoionization cross section
of one particular excited level $2s^22p^5[^2P^o_{3/2}]4d(^1F^o_3)$
(ionization energy =
17.626 Ry) of \fexvii at four
densities.  The main features are as follows: (I) orders of magnitude
variation in resonance heights and extent. (II) 
For $N_e > 10^{20}$/cc AI resonances begin to exhibit 
broadening and smearing of overlapping Rydberg series.
$N_e = 10^{21}$cc. The narrower high-\en \el resonances dissolve into
the
continua but stronger low-\en \el resonance retain their characteristic
asymmetric shapes. With increasing density
$N_e=10^{21-23}$cc, resonance
structures not only broaden but their strengths shift and redistributed
over a wide range determined by the total width
$\Gamma(\omega,\nu_i,T,N_e)$ at each photon energy $\hbar \omega$
(Eq.~\ref{eq:sig}).
(III) The averaged step-wise structure due to 
Stark ionization cut-off (Table~1)
represents complete dissolution into continua.
(IV) It is numerically ascertained that
total AI resonance strengths are conserved, and
integrated values generally do not deviate by more than 1-2\%.
This is also an important accuracy
check on numerical integration and the computational algorithm,
as well as the choice of the parameter $\delta$ that determines the
energy range of the Lorentizan profile at each T and $N_e$. In the
present $\delta$ = 0.01 for all \dne=$10^{20-23}$/cc.

Plasma effects on AI features Fig.~\ref{fig:f1}
show a redward shift of low-\en resonances and dissolution of high-\en
resonances. In addition, the background continuum is raised owing to
redistribution of resonance strengths, which merge into one across high
lying and overlapping thresholds. 
The shifts in AI resonance strengths, akin to line shifts but much more
pronouned, is particularly important since cross sections are integrated
over plasma particle distributions in order to obtain rates for atomic
proceses. Also noteworthy is the height and extent of prominent
resonances features dominated by Seaton PEC resonances. cross sections
may range up to 10 orders magnitude in height and hundreds of eV in
energy.

 Table~1 gives plasma parameters corresponding to \fexviii at
along two plasma isotherms and varying densities.
The maximum
width $\Gamma_{10}$ corresponding to $\nu_i=10$ in
Eqs.~\ref{eq:gam},\ref{eq:gamc},\ref{eq:gams} 
corresponding to the $\nu$-mesh at which
the unbroadened AI resonance profiles are delineated
up to $\nu \leq 10$; an averaging procedure is employed up to
$ 10 < \nu < \infty$ using quantum defect (QD)
theory (viz. \cite{op,aas}).
$\Gamma_c(10)$ and $\Gamma_s(10)$ are the maximum collisional and Stark
width components. The thermal Doppler width $\Gamma_d$ is much smaller,
as may be inferred from the fact that the total width \gam(10)
$\approx$ \gamc + \gams. However, in lower density plasmas $N_e <
10^{20}$/cc, $\Gamma_d$ may be
comparable to $\Gamma_c$ or $\Gamma_s$. 

In Table~1, the $\nu^*_s$ and
$\nu_D$ are effective quantum numbers corresponding to Stark
ionization cut-off and the Debye radius respectively. For HED plasmas
with $N_e > 10^{23}$, one needs to examine if the bound orbitals are
penetrated by the free electrons as the Debye length increases, and
plasma screening effects may need to be considered.
We therefore calculate the corresponding effective quantum number
$\nu_D = \left[ \frac{2}{5}\pi z^2 \lambda_D^2 \right ]^{1/4}$,
where the Debye length $\lambda_D = (kT/8\pi N_e)^{1/2}$.
It is seen in Table~1 that
$\nu_D > \nu^*_s$ for all T, \dne considered, justifying neglect of
plasma screening effects herein,
but which may need to be accounted for at even higher
densities.

AI broadening
in a plasma environment
affects each level cross section differently, and hence
its contribution to opacities or rate equations for atomic processes
in general. A critical (T,\dne) range can therefore be numerically
ascertained where redistributed resonance phenomena
 would be significant and cross sections should be modified.

\begin{table}
\caption{Plasma parameters along isotherms $10^6$K and $2\times10^6$K,
and electron densities $N_e = 10^{20-23}$/cc
for the (e~+~\fexix) $\longrightarrow$ \fexviii system w.r.t
Fig.~\ref{fig:f1}. Total
AI resonance widths are shown at $\nu \approx 10$,
with corresponding collisional widths \gamc(10) and Stark width \gams(10).
Effective quantum number $\nu_s^{\ast}$ refers to Stark ionization, and 
$\nu_D$ corresponds to the Debye radius. AI broadening widths are very
weakly dependent on temperature and thermal Doppler widths are
negligible in comparison with \gamc and \gams.}
\begin{center}
\begin{tabular} {c|c|c|c|c|c|c}
\hline
 T(K) & $N_e (cc)$ & $\Gamma(\nu=10)$ & $\Gamma_c(10)$ & $\Gamma_s(10)$
& $\nu_s^*$ & $\nu_D$ \\
& & Ry & Ry & Ry & &\\
\hline
$10^6$ & $10^{20}$ & 5.98(-2)& 7.56(-3) & 5.23(-2) & 14.6 & 43.2\\
$10^6$ & $10^{21}$ & 3.18(-1)& 7.56(-2) & 2.43(-2) & 10.8 & 24.3\\
$10^6$ & $10^{22}$ & 1.88(0)& 7.56(-1) & 1.13(0) & 7.93 & 13.7\\
$10^6$ & $10^{23}$ & 1.28(1)& 7.56(0) & 5.22(0) & 5.83 & 7.68\\
\hline
$2\times10^6$ & $10^{20}$ & 5.97(-2)& 7.45(-3) & 5.23(-2) & 14.6 &
51.4\\
$2\times10^6$ &  $10^{21}$ & 3.17(-1) & 7.45(-2) & 2.43(-1) & 10.8 &
28.9  \\
$2\times10^6$ &  $10^{22}$ & 1.87(0) & 7.45(-1) & 1.13(0) & 7.93 & 16.2
 \\
$2\times10^6$ &  $10^{23}$ & 1.27(1) & 7.45(0) & 5.23(0) & 5.83 & 9.13
\\
\hline
\end{tabular}
\end{center}
\end{table}

\section{Conclusion}

 Atomic cross sections and rates in 
HED plasma sources at sufficiently high densities may be significantly
affected by attenuation and broadening of AI resonant features.
Precise evaluation of equation-of-state of the plasma determines the
number of levels in predicting macroscopic properties such as opacity
and radiation transport.
 Utilizing the R-matrix framwork, a computationally viable theoretical tratment 
taking account of plasma effects is
reviewed. The method generalizes the description of AI phenomena
of isolated Fano profiles in plasmas. Analogous to line shapes,
atomic cross sections with resonant features
become energy-temperature-density dependent, leading to
broadening, shifting, and dissolving into myriad \eion
continua. However, unlike symmetric line profiles, 
the intrinsically asymmetric AI resonance shapes are attenuated over
extended energy ranges.
The predicted energy shifts of AI resonances as the plasma
density increases should be experimentally verifiable.
Redistribution of AI resonance strengths should 
manifest itself in rate coefficients for
\eion excitation, \eion recombination, photoionization, opacities and
radiation transport in HED plasma models,
using temperature-dependent
Maxwellian, Planck, or other particle distribution functions.
The computational algorithm and a general-purpose program has been
developed for large-scale computations.
Finally, the complex electron correlation and coupling effects may now
be computed precisely for myriad HED plasma applications using the 
R-matrix method.

\vskip 0.25in
{\bf Acknowledgments}
\vskip 0.25in

 I would like to thank Sultana Nahar for atomic data for Fe ions used in
plasma AI broadening calculations.

\end{document}